\documentclass[12pt,twoside]{article}

\usepackage{pslatex}	
\usepackage{amsfonts}
\usepackage{amsmath}
\usepackage{amsthm}
\usepackage{amscd}
\usepackage{cite}
\usepackage{epsf}
\usepackage{a4}

\def\beq{\begin{equation}}
\def\eeq{\end{equation}}
\def\bea{\begin{eqnarray}}
\def\eea{\end{eqnarray}}

\def\beann{\begin{eqnarray*}}
\def\eeann{\end{eqnarray*}}

\let\a=\alpha \let\be=\beta \let\g=\gamma \let\de=\delta

  \let\la=\lambda \let\m=\mu
  \let\p=\pi  \let\s=\sigma
 
\let\ph=\varphi  \let\PH=\Phi 
\let\Om=\Omega  
\let\La=\Lambda  

\let\qd=\quad \let\qqd=\qquad 

\def\epp{\, .}
\def\epc{\, ,}

\def\tst#1{{\textstyle #1}}

\theoremstyle{plain}

\newtheorem{lemma}{Lemma}

\newtheorem*{corollary*}{Corollary}

\theoremstyle{definition}

\def\2{\frac{1}{2}} \def\4{\frac{1}{4}}

\def\6{\partial}

\def\+{\dagger}

\def\<{\langle} \def\>{\rangle}

\let\auf=\uparrow \let\ab=\downarrow

\def\CH{{\cal H}} \def\CL{{\cal L}}
 
  \def\CT{{\cal T}}

\def\i{{\rm i}}

 \def\ch{{\rm ch}}

 \def\End{{\rm End}} \def\id{{\rm id}}

\def\diag{{\rm diag}}

\def\str{{\rm str}}

\def\xv{\mathbf{x}}
\def\yv{\mathbf{y}}

\renewcommand{\appendix}{%
   \renewcommand{\section}{
	\secdef\Appendix\sAppendix}%
   \setcounter{section}{0}%
   \renewcommand{\thesection}{\Alph{section}}%
   \renewcommand{\theequation}{\thesection.\arabic{equation}}%
}

\newcommand{\Appendix}[2][?]{%
     \refstepcounter{section}%
     \setcounter{equation}{0}%
     \addcontentsline{toc}{appendix}%
          {\protect\numberline{\appendixname~\thesection} #1}%
     \vspace{\baselineskip}%
     {\noindent\large\bfseries\appendixname\ \thesection: #2\par}%
     \sectionmark{#1}\vspace{\baselineskip}}

\newcommand{\sAppendix}[1]{%
     {\noindent\large\bfseries\appendixname\: #1\par}%
     \sectionmark{#1}\vspace{\baselineskip}}

\begin{document}

\thispagestyle{empty}

\begin{center}

{\Large {\bf Algebraic Bethe ansatz for the gl(1$|$2) generalized\\
model II: the three gradings\\}}

\vspace{7mm}

{\large Frank G\"{o}hmann\footnote[2]{e-mail:
goehmann@physik.uni-wuppertal.de} and Alexander Seel\footnote{e-mail:
seel@physik.uni-wuppertal.de}\\

\vspace{5mm}

Fachbereich Physik, Bergische Universit\"at Wuppertal,\\
42097 Wuppertal, Germany\\}

\vspace{20mm}

{\large {\bf Abstract}}

\end{center}

\begin{list}{}{\addtolength{\rightmargin}{10mm}
               \addtolength{\topsep}{-5mm}}
\item
The algebraic Bethe ansatz can be performed rather abstractly for
whole classes of models sharing the same $R$-matrix, the only
prerequisite being the existence of an appropriate pseudo vacuum
state. Here we perform the algebraic Bethe ansatz for all models
with $9 \times 9$, rational, gl(1$|$2)-invariant $R$-matrix and all
three possibilities of choosing the grading. Our Bethe ansatz solution
applies, for instance, to the supersymmetric $t$-$J$ model, the
supersymmetric $U$ model and a number of interesting impurity models.
It may be extended to obtain the quantum transfer matrix spectrum
for this class of models. The properties of a specific model enter
the Bethe ansatz solution (i.e.\ the expression for the transfer
matrix eigenvalue and the Bethe ansatz equations) through the three
pseudo vacuum eigenvalues of the diagonal elements of the monodromy
matrix which in this context are called the parameters of the model.
\\[2ex]
{\it PACS: 05.50.+q, 71.10.Fd, 71.10.Pm}
\end{list}

\clearpage

\section{Introduction}
This work resumes previous work of one of the authors \cite{Goehmann02}
where the algebraic Bethe ansatz for the gl(1$|$2) generalized
model was constructed for the grading $(+,-,-)$. In this article we
address the two remaining cases $(-,+,-)$ and $(-,-,+)$ which turned
out to be technically more involved, since the grading enters the
auxiliary second level Bethe ansatz in a non-trivial way (see
appendix~\ref{app:derev}).

Performing an algebraic Bethe ansatz calculation means to diagonalize
the transfer matrix of a certain two-dimensional classical vertex model
by purely algebraic means or (in a physicists language) by using only
commutation relations between operators. If the transfer matrix has a
Hamiltonian limit this is equivalent to diagonalizing the Hamiltonian
along with its conserved currents. The Hamiltonian and the conserved
currents are then usually generated by expanding the logarithm of the
transfer matrix in the spectral parameter.

The algebraic Bethe ansatz can be performed on a considerable level
of abstraction and seems to depend only on the structure of the
$R$-matrix of a given model and on the existence of a so-called pseudo
vacuum or highest vector \cite{KuSk82b} on which the monodromy matrix
acts as an (upper) triangular matrix. This idea was first of
all worked out for the models falling into the same class as the
XXZ spin chain \cite{Korepin82b} and turned out to be useful in
calculating the norm \cite{Korepin82} and certain matrix elements
\cite{Slavnov89} of Bethe ansatz states. In \cite{Korepin82,Korepin82b}
V.~E.~Korepin  introduced the notion of a `generalized model' whose
`representation' is given by the action of the diagonal elements
of the monodromy matrix on the pseudo vacuum. He assumed the vacuum
eigenvalues, say $a_1 (\la)$ and $a_2 (\la)$, to be arbitrary and
called them (functional) parameters of the model. Later
\cite{Tarasov84,Tarasov85} V.~O.~Tarasov refined and basically
confirmed Korepin's concept.

The algebraic Bethe ansatz for the generalized model associated with
the $R$-matrix of the XXZ model is of the same structure as for the
fundamental XXZ spin chain. Such simple relation holds no longer
for models of `higher rank' which require a nested Bethe ansatz.
The simplest models which allow for a nested algebraic Bethe
ansatz are the models with gl(n) invariant $R$-matrix \cite{KuSk80,%
KuRe82}. Considering the fundamental representations of these models
one observes that not only the monodromy matrix elements below the
diagonal annihilate the pseudo vacuum, but additional zeros appear
above the diagonal \cite{KuRe81}. This fact simplifies the algebraic
Bethe ansatz for the fundamental representation as compared to the more
general case, where the action of all the elements of the monodromy
matrix above the diagonal is non-trivial. For the solution of this more
general case a new concept, the vacuum subspace, was introduced by
Kulish and Reshetikhin \cite{KuRe83}. This new concept made it possible
to perform the algebraic Bethe ansatz for the models with gl(n)
invariant $R$-matrix on the same level of generality as in the gl(2)
case (corresponding to the $R$-matrix of the XXX spin chain of spin~%
$\2$). The resulting eigenvalue of the transfer matrix and the Bethe
ansatz equations depend on $n$ functional parameters $a_1 (\la), \dots,
a_n (\la)$, which, together with the triangular action of the monodromy
matrix on the pseudo vacuum, define the gl(n) generalized model
\cite{Reshetikhin86}. Considering the parameters $a_1 (\la), a_2 (\la),
a_3 (\la)$ as free Reshetikhin derived the norm formula for the
gl(3) case \cite{Reshetikhin86}.

To our knowledge there was no more activity in the direction of
constructing algebraic Bethe ansatz solutions of generalized models
beyond the above mentioned work of Kulish and Reshetikhin. This may
partially be related to the general difficulties in generalizing the
algebraic Bethe ansatz beyond gl(n) (see e.g. \cite{MaRa97}). For the
models with gl(1$|$2)-invariant $R$-matrix (e.g.\ \cite{Lai74,%
Sutherland75,Schlottmann87,Karnaukhov94,BGLZ95,Maassarani95,BEF97,%
Frahm99,FLT99}\footnote{For a more thorough discussion see the
introduction of \cite{Goehmann02}.}), which (together with its
anisotropic generalizations) are of particular interest for application
in solid state physics, many algebraic Bethe ansatz solutions were
constructed \cite{Kulish85,EsKo92,FoKa93,RaMa96,HGL96,PfFr96,PfFr97,%
BEF97,Frahm99} which can all (except for \cite{RaMa96}) be interpreted
as certain realizations of the solution obtained in \cite{Goehmann02}
and are extended here to the two gradings $(-,+,-)$ and $(-,-,+)$ not
treated in \cite{Goehmann02}.

Future applications of our work may be the quantum transfer matrix
approach to the thermodynamics of the models with gl(1$|$2) invariant
$R$-matrix \cite{JKS97,KWZ97}, the calculation of norms and matrix
elements and, possibly (see \cite{Goehmann02}), the algebraic Bethe
ansatz for the Hubbard model.

Our article is organized as follows. Section 2 contains the
basic definitions relating to the gl(1$|$2) generalized model with
some material shifted to appendix \ref{app:defga}. In section~3 we
discuss how the grading and the Yang-Baxter algebra change under
permutations of the basis vectors in auxiliary space. We shall see
how the grading may change with the change of the highest vector.
Section 4 contains our main result, the formulae for the algebraic
Bethe ansatz solution of the generalized model for the three possible
gradings. For the grading $(+,-,-)$ a derivation was presented in
\cite{Goehmann02}. The proof for the remaining two cases is sketched
in appendix \ref{app:derev}. Section 5 contains examples of how to
apply the formulae of section 4. We mainly reconsider the well known
example of the fundamental supersymmetric $t$-$J$ model \cite{EsKo92,%
FoKa93} and clarify the connection between the possible choices of
pseudo vacua and possible gradings. We also obtained a simple proof of
the equivalence of the different Bethe ansatz solutions of the
supersymmetric $t$-$J$ model (which exceeds the one in \cite{EsKo92},
since we also show that the eigenvalues are identical). This proof
\cite{GoSe03b} will be published separately. Section 6 contains our
conclusion and a discussion of future perspectives.

\section{The gl(1$|$2) generalized model}
We begin by specifying the class of models we are going to consider.
As was explained in the introduction this class is determined by its
$R$-matrix and by the existence of a pseudo vacuum. Here the $R$-matrix
has matrix elements of the form
\begin{equation} \label{rm}
     R^{\a \g}_{\be \de} (\la) = a(\la) (-1)^{p(\a) p(\g)}
                                 \de^\a_\be \de^\g_\de +
				 b(\la) \de^\a_\de \de^\g_\be \epc
\end{equation}
$\a, \be, \g, \de = 1, 2, 3$. This $R$-matrix is based on graded
permutations \cite{Sutherland75}. It is contained in the early list
of Kulish and Sklyanin \cite{KuSk80}. The algebraic Bethe ansatz
for the fundamental model was constructed by Kulish in 1985
\cite{Kulish85} along with the general gl(m$|$n) case.

The $R$-matrix (\ref{rm}) is characterized by two rational, complex
valued functions
\begin{equation} \label{defab}
     a(\la) = \frac{\la}{\la + \i c} \epc \qd
     b(\la) = \frac{\i c}{\la + \i c}
\end{equation}
depending on a spectral parameter $\la \in {\mathbb C}$ and a coupling
$c \in {\mathbb C}$. It further depends on the grading, $p:\{1,2,3\}
\rightarrow {\mathbb Z}_2$. We shall consider three cases:%
\footnote{We comment on the remaining possibilities of choosing the
grading below in section \ref{sec:vargrad}.}
\begin{equation}
\label{thethree}
\begin{aligned}
     {\rm (i)} & & p(1) = 0 \epc & & p(2) = p(3) = 1 \epc \\
     {\rm (ii)} & & p(2) = 0 \epc & & p(3) = p(1) = 1 \epc \\
     {\rm (iii)} & & p(3) = 0 \epc & & p(1) = p(2) = 1 \epp
\end{aligned}
\end{equation}
In order to refer to these different cases we introduce a vector index
$g$ which is $g = (g_1, g_2, g_3) = (+,-,-)$ in the first case,
$g = (-,+,-)$ in the second case, and $g = (-,-,+)$ in the third case.
We shall say `the grading is $g$', and we shall write $R_g$ instead
of $R$.

The matrix $R_g(\la)$ solves the Yang-Baxter equation and obviously
satisfies the compatibility condition \cite{KuSk80}
\begin{equation} \label{comp}
     {R_g}^{\a \g}_{\be \de} (\la) =
        (-1)^{p(\a) + p(\be) + p(\g) + p(\de)}
	{R_g}^{\a \g}_{\be \de} (\la) \epp
\end{equation}
In order to introduce the notion of the graded Yang-Baxter algebra
we shall further need the matrix $\check R_g (\la)$ defined by
switching the row indices of $R_g (\la)$,
\begin{equation} \label{defrcheck}
     \check {R_g}^{\a \g}_{\be \de} (\la) =
        {R_g}^{\g \a}_{\be \de} (\la) \epp
\end{equation}

The graded Yang-Baxter algebra with $R$-matrix $R(\la)$ is the graded,
associative algebra (with unity) generated by the elements
$\CT^\a_\be (\la)$, $\a, \be = 1, 2, 3$, of the so-called monodromy
matrix modulo the relations
\begin{equation} \label{gyba}
     \check R (\la - \m) \big( \CT (\la) \otimes_g \CT (\m) \big) =
     \big( \CT (\m) \otimes_g \CT (\la) \big) \check R (\la - \m) \epp
\end{equation}
We shall assume that the elements of the monodromy matrix are of
definite parity, $\p(\CT^\a_\be (\la)) = p(\a) + p(\be)$. The symbol
$\otimes_g$ denotes the super tensor product associated with the
grading $g$. For a definition see appendix~A.

We now define the set of all models related to the $R$-matrix (\ref{rm})
and solvable by algebraic Bethe ansatz as `the gl(1$|$2) generalized
model': By definition the gl(1$|$2) generalized model is the set of all
(linear) representations of the graded Yang-Baxter algebra (\ref{gyba})
having a highest vector (or pseudo vacuum) $\Om$. The highest vector
$\Om$ is a vector on which the monodromy matrix $\CT (\la)$ acts as an
upper triangular matrix and which is an eigenvector of its diagonal
elements:
\begin{equation} \label{para}
\begin{split}
     & \CT^1_1 (\la) \Om = a_1 (\la) \Om \epc \qd
       \CT^2_2 (\la) \Om = a_2 (\la) \Om \epc \qd
       \CT^3_3 (\la) \Om = a_3 (\la) \Om \, , \\[1ex]
     & \CT^\a_\be (\la) \Om = 0 \epc \qd \text{for $\a > \be$} \epp
\end{split}
\end{equation}
The eigenvalues $a_j (\la)$, $j = 1, 2, 3$ of the diagonal elements of
$\CT(\la)$, are called the parameters of the generalized model. These
parameters characterize the representation in a similar manner as the
highest weight in a highest weight representation of a Lie algebra.

Let us denote the representation space of a given representation of the
generalized model by $\CH$. It is clear from the quadratic commutation
relations contained in the graded Yang-Baxter algebra (\ref{gyba}) and
from (\ref{para}) that we may assume $\CH$ to be spanned by all
vectors of the form
\begin{equation}
     \PH (\la_1, \dots, \la_N) = \CT^{\a_1}_{\be_1} (\la_1) \dots
                                 \CT^{\a_N}_{\be_N} (\la_N) \Om \epp
\end{equation}
where $\a_k < \be_k$, $k = 1, \dots, N$. This assumption is at least
sensible for a finite dimensional representation space $\CH$.

The super trace of the monodromy matrix
\begin{equation} \label{deftr}
     t(\la) = (-1)^{p(\a)} \CT^\a_\a (\la) = \str_g (\CT (\la)) \epp
\end{equation}
is called the transfer matrix of the generalized model. Since $\check
R (\la)$ is invertible for generic values of $\la \in {\mathbb C}$, we
conclude from (\ref{comp}) and (\ref{gyba}) that the transfer matrix
satisfies
\begin{equation}
     [t(\la), t(\m)] = 0
\end{equation}
for all generic $\la, \m \in {\mathbb C}$. It follows that $t(\la)$ and
$t(\m)$ have a common system of eigenfunctions, which means that the
eigenvectors of $t(\la)$ are independent of the spectral parameter
$\la$.

The task to be performed below of the algebraic Bethe ansatz for the
generalized model is to diagonalize $t(\la$), i.e., to solve the
eigenvalue problem
\begin{equation} \label{trmevp}
     t(\la) \PH = \La (\la) \PH \epp
\end{equation}
It is a remarkable fact that this task can be accomplished by solely
using the graded Yang-Baxter algebra (\ref{gyba}) and the properties
(\ref{para}) of the highest vector $\Om$. In particular, it is
{\em not} necessary to require that $\CT^2_3 (\la) \Om = 0$ as in case
of the fundamental graded representation, which corresponds to the
supersymmetric $t$-$J$ model.

\section{Variation of the grading}
\label{sec:vargrad}
Before presenting our results for the algebraic Bethe ansatz we would
like to explain why we may restrict ourselves to upper triangular
action in our definition (\ref{para}) of the highest vector $\Om$ and
why we consider only the three gradings shown in equation
(\ref{thethree}). For the former purpose we first of all introduce the
natural action of the symmetric group ${\mathfrak S}^3$ on row vectors
$\xv = (x_1, x_2, x_3)$, setting
\begin{equation}
     \xv \s = (x_{\s(1)}, x_{\s(2)}, x_{\s(3)}) \epc
\end{equation}
for all $\s \in {\mathfrak S}^3$. This defines a faithful representation
of ${\mathfrak S}^3$ which is orthogonal with respect to the usual
Euclidian scalar product $\< \xv, \yv\> = x_1 y_1 + x_2 y_2 + x_3 y_3$.

Denoting the transposed of $\s$ by $\s^t$ we obtain the transformation
properties of the $R$-matrix under permutations directly from its
definition (\ref{rm}),
\begin{equation}
     (\s^t \otimes \s^t) R_g (\la) (\s \otimes \s) =
        R_{g \s} (\la) \epp
\end{equation}
Similarly, the graded Yang-Baxter algebra (\ref{gyba}) is easily seen
to transform as
\begin{equation} \label{permgyba}
     \check R_{g \s} (\la - \m) \bigl( \s^t \CT (\la) \s \otimes_{g \s}
        \s^t \CT (\m) \s \bigr) =
     \bigl( \s^t \CT (\m) \s \otimes_{g \s} \s^t \CT (\la) \s \bigr)
        \check R_{g \s} (\la - \m) \epc
\end{equation}
while it follows from (\ref{deftr}) that
\begin{equation} \label{tperminv}
     \str_g (\CT(\la)) = \str_{g \s} \bigl( \s^t \CT (\la) \s \bigr)
                            \epc
\end{equation}
which expresses the invariance of the transfer matrix with respect to
permutations.

\begin{figure}
\[
\begin{CD}
     \begin{pmatrix} a_1 & \ast & \ast \\ 0 & a_2 & \ast \\
                     0 & 0 & a_3 \end{pmatrix} \Om
     @>{\Pi_{12}}>>
     \begin{pmatrix} a_2 & 0 & \ast \\ \ast & a_1 & \ast \\
                     0 & 0 & a_3 \end{pmatrix} \Om
     @>{\Pi_{23}}>>
     \begin{pmatrix} a_2 & \ast & 0 \\ 0 & a_3 & 0 \\
                     \ast & \ast & a_1 \end{pmatrix} \Om
     \\[2ex] @A{\Pi_{23}}AA @. @VV{\Pi_{12}}V\\[2ex]
     \begin{pmatrix} a_1 & \ast & \ast \\ 0 & a_3 & 0 \\
                     0 & \ast & a_2 \end{pmatrix} \Om
     @<<{\Pi_{12}}<
     \begin{pmatrix} a_3 & 0 & 0 \\ \ast & a_1 & \ast \\
                     \ast & 0 & a_2 \end{pmatrix} \Om
     @<<{\Pi_{23}}<
     \begin{pmatrix} a_3 & 0 & 0 \\ \ast & a_2 & 0 \\
                     \ast & \ast & a_1 \end{pmatrix} \Om
\end{CD}
\]
\caption{Change of the monodromy matrix action on the highest vector
$\Om$ under permutations of the basis vectors in auxiliary space.}
\label{fig:permon}
\end{figure}

In figure \ref{fig:permon} we show the action of the transformed
monodromy matrix $\s^t \CT (\la) \s$ on a highest vector $\Om$, when
$\s$ runs successively through all permutations in ${\mathfrak S}^3$
generated by the transpositions of nearest neighbours $\Pi_{12}$ and
$\Pi_{23}$. We see that if, for a given grading $g$, a monodromy
matrix $\CT (\la)$ realizes one of the six patterns in figure
\ref{fig:permon} by acting on some vector $\Om$ then there is a
permutation $\s \in {\mathfrak S}^3$ such that $\s^t \CT(\la) \s$ acts
as an upper triangular matrix on $\Om$. The corresponding grading
changes from $g$ to~$g \s$.

Let us consider an example. Take $g = (-,-,+)$ and $\CT (\la)$ and $\Om$
such that
\begin{equation}
     \CT (\la) \Om =
     \begin{pmatrix} a_1 & 0 & 0 \\ \ast & a_2 & \ast \\
                     \ast & 0 & a_3 \end{pmatrix} \Om \epp
\end{equation}
This is the pattern in the middle of the second row in figure
\ref{fig:permon}. Thus,
\begin{equation}
     \Pi_{23} \Pi_{12} \CT (\la) \Pi_{12} \Pi_{23} \; \Om = 
     \begin{pmatrix} a_2 & \ast & \ast \\ 0 & a_3 & \ast \\
                     0 & 0 & a_1 \end{pmatrix} \Om \epc
\end{equation}
and the grading changes to $(-,-,+) \Pi_{12} \Pi_{23} = (-,+,-)$.

Combining the six patterns in figure \ref{fig:permon} with the three
gradings $g = (+,-,-)$, $(-,+,-)$, $(-,-,+)$ we obtain 18 cases which by
application of permutations all reduce back to three, e.g.\ upper
triangular action with three different gradings. The Bethe ansatz
solutions of the transfer matrix eigenvalue problem (\ref{trmevp}) for
these three cases will be presented in the next section.

Note that it may happen that there are several vectors $\Om_1, \Om_2,
\dots$ which for given monodromy matrix and grading generate several
of the patterns in figure (\ref{fig:permon}). Then there are
several equivalent but differently looking Bethe ansatz solutions of
the transfer matrix eigenvalue problem. This is, for instance, the
case for the supersymmetric $t$-$J$ model as was observed in
\cite{EsKo92}. We will come back to this phenomenon in our example
section (see section \ref{sec:examples}).

How about the other possible gradings? There are eight cases
altogether. Three cases are listed in (\ref{thethree}). The case
$g = (+,+,+)$ was treated by Kulish and Reshetikhin \cite{KuRe83}. The
four remaining cases are related to our work or the work of Kulish and
Reshetikhin by a switch of sign of $g$ (e.g.\ $(+,-,-) \rightarrow
(-,+,+)$). It is easy to see that this switch modifies the Yang-Baxter
algebra (\ref{gyba}) only trivially: We introduce a diagonal matrix
$G = \diag(g_1, g_2, g_3)$ and the $3 \times 3$ unit matrix $I_3$. Then
a similarity transformation with $G \otimes I_3$ transforms the graded
Yang-Baxter algebra (\ref{gyba}) with $R$-matrix $R_g$ into
\begin{equation} \label{minusgyba}
     \check R_{-g} (\la - \m)
        \big( \CT (- \la) \otimes_{-g} \CT (- \m) \big) =
        \big( \CT (- \m) \otimes_{-g} \CT (- \la) \big)
	\check R_{-g} (\la - \m) \epp
\end{equation}
Note that the expression for the parity of the monodromy matrix
elements, $\p(\CT^\a_\be (\la)) = p(\a) + p(\be)$, is invariant under
a change of the sign of $g$, since it corresponds to replacing
$p(\a)$ by $p(\a) + 1$. Thus, every representation with parameters
$a_j (\la)$ of the graded Yang-Baxter algebra (\ref{gyba}) with
grading $g$ and $R$-matrix $R_g (\la)$ is equivalent to a
representation with parameters $a_j (- \la)$ of (\ref{gyba}) with
grading $-g$ and $R$-matrix $R_{-g} (\la)$. Consequently the Bethe
ansatz solutions of the generalized model for the remaining gradings
are obtained from the solutions in the following section by switching
the sign in the argument of $a(\la)$ and the overall signs of the
transfer matrix eigenvalues (or by performing similar manipulations
in the corresponding equations in \cite{KuRe83}).

\section{The algebraic Bethe ansatz solution} \label{sec:res}
In this section we list the transfer matrix eigenvalues and the
corresponding eigenvectors. For $g = (+,-,-)$ they were obtained in
\cite{Goehmann02}. The derivations for the two remaining cases
$g = (-,+,-)$ and $g = (-,-,+)$ are presented in appendix
\ref{app:derev}.

Recall that the functions $a_1$, $a_2$ and $a_3$ are the (functional)
parameters of the model and that $a(\la) = \la/(\la + \i c)$. The
different transfer matrix eigenvalues $\La_g (\la)$ (equations
(\ref{lapmm}), (\ref{lampm}), (\ref{lammp}) below) are distinguished
for a given grading by specifying two sets of Bethe roots
$\{\la_j\}_{j=1}^N$ and $\{\m_k\}_{k=1}^M$ which have to be calculated
from two coupled sets of Bethe ansatz equations (see (\ref{bapmm}),
(\ref{bampm}), (\ref{bammp}) below).
\begin{align} \label{lapmm}
     \La_{(+--)} (\la) =
                   & a_1 (\la) \prod_{j=1}^N \frac{1}{a(\la_j - \la)}
                      \notag \\[-1ex]
                   & \: - a_2 (\la) \prod_{j=1}^N
		                     \frac{1}{a(\la_j - \la)}
		                     \prod_{k=1}^M
				     \frac{1}{a(\la - \m_k)}
			 - a_3 (\la) \prod_{k=1}^M
			             \frac{1}{a(\m_k - \la)} \epc
\end{align}
\begin{subequations}
\label{bapmm}
\begin{align}
     \frac{a_1 (\la_j)}{a_2 (\la_j)} & = \prod_{k=1}^M
                                         \frac{1}{a(\la_j - \m_k)} \epc
			 		 \qd j = 1, \dots, N \epc \\
     \frac{a_3 (\m_k)}{a_2 (\m_k)} & =
          \prod_{\substack{l = 1 \\ l \ne k}}^M
	  \frac{a(\m_l - \m_k)}{a(\m_k - \m_l)}
	  \prod_{j=1}^N \frac{1}{a(\la_j - \m_k)} \epc
	                                 \qd k = 1, \dots, M \epp
\end{align}
\end{subequations}
\begin{align} \label{lampm}
     \La_{(-+-)} (\la) =
                   & - a_1 (\la) \prod_{j=1}^N \frac{1}{a(\la - \la_j)}
                      \notag \\[-1ex]
                   & \: + a_2 (\la) \prod_{j=1}^N
		                     \frac{1}{a(\la - \la_j)}
		                     \prod_{k=1}^M
				     \frac{1}{a(\m_k - \la)}
			 - a_3 (\la) \prod_{k=1}^M
			             \frac{1}{a(\m_k - \la)} \epc
\end{align}
\begin{subequations}
\label{bampm}
\begin{align}
     \frac{a_1 (\la_j)}{a_2 (\la_j)} & = \prod_{k=1}^M
                                         \frac{1}{a(\m_k - \la_j)} \epc
			 		 \qd j = 1, \dots, N \epc \\
     \frac{a_3 (\m_k)}{a_2 (\m_k)} & = \prod_{j=1}^N
                                       \frac{1}{a(\m_k - \la_j)} \epc
	                               \qd k = 1, \dots, M \epp
\end{align}
\end{subequations}
\begin{align} \label{lammp}
     \La_{(--+)} (\la) =
                   & - a_1 (\la) \prod_{j=1}^N \frac{1}{a(\la - \la_j)}
                      \notag \\[-1ex]
                   & \: - a_2 (\la) \prod_{j=1}^N
		                     \frac{1}{a(\la_j - \la)}
		                     \prod_{k=1}^M
				     \frac{1}{a(\la - \m_k)}
			 + a_3 (\la) \prod_{k=1}^M
			             \frac{1}{a(\la - \m_k)} \epc
\end{align}
\begin{subequations}
\label{bammp}
\begin{align}
     \frac{a_1 (\la_j)}{a_2 (\la_j)} & =
          \prod_{\substack{l = 1 \\ l \ne j}}^N
	  \frac{a(\la_j - \la_l)}{a(\la_l - \la_j)}
	  \prod_{k=1}^M \frac{1}{a(\la_j - \m_k)} \epc
			 		 \qd j = 1, \dots, N \epc \\
     \frac{a_3 (\m_k)}{a_2 (\m_k)} & = \prod_{j=1}^N
                                       \frac{1}{a(\la_j - \m_k)} \epc
	                               \qd k = 1, \dots, M \epp
\end{align}
\end{subequations}
These three sets of expressions for the eigenvalues and Bethe ansatz
equations depend on the grading $g = (g_1, g_2, g_3)$ in a
characteristic way which allows us two write them all in one
(for a similarly compact expression for the ($q$-deformed) fundamental
model see \cite{Schultz83}):
\begin{multline} \label{laggg}
     \La_g (\la) =
        g_1 a_1 (\la) \prod_{j=1}^N \frac{1}{a(g_1 (\la_j - \la))}
	   \\[-1ex] + g_2 a_2 (\la)
	       \prod_{j=1}^N \frac{1}{a(g_2 (\la - \la_j))}
	       \prod_{k=1}^M \frac{1}{a(g_2 (\m_k - \la))}
             + g_3 a_3 (\la)
	       \prod_{k=1}^M \frac{1}{a(g_3 (\la - \m_k))} \epc
\raisetag{12ex}
\end{multline}
\begin{subequations}
\label{baggg}
\begin{align}
     \frac{a_1 (\la_j)}{a_2 (\la_j)} & =
          \prod_{\substack{l = 1 \\ l \ne j}}^N
	  \frac{a(g_1 (\la_l - \la_j))}{a(g_2 (\la_j - \la_l))}
	  \prod_{k=1}^M \frac{1}{a(g_2 (\m_k - \la_j))} \epc
			 		 \qd j = 1, \dots, N \epc \\
     \frac{a_3 (\m_k)}{a_2 (\m_k)} & =
          \prod_{\substack{l = 1 \\ l \ne k}}^M
	  \frac{a(g_3 (\m_k - \m_l))}{a(g_2 (\m_l - \m_k))}
	  \prod_{j=1}^N \frac{1}{a(g_2 (\m_k - \la_j))} \epc
	                                 \qd k = 1, \dots, M \epp
\end{align}
\end{subequations}
Our notation means, for instance, that by specifying $g_1 = - 1$,
$g_2 = 1$, $g_3 = - 1$ equations (\ref{laggg}), (\ref{baggg}) turn
into (\ref{lampm}), (\ref{bampm}) corresponding to the grading
$g = (-,+,-)$.

Describing the corresponding eigenvectors requires more effort,
since we will have to introduce several notions related to the
`second Bethe ansatz' in the nested Bethe ansatz calculation that led
to the above expressions for the eigenvalues. The eigenvectors are
obtained by acting with certain linear combinations of products of
monodromy matrix elements on the highest vector $\Om$. They are of
the form
\begin{equation} \label{evform}
     \PH_g (\la_1, \dots \la_N; \m_1, \dots, \m_M) =
        \sum_{i_1, \dots, i_N = 1}^2
           B_{i_1}^g (\la_1) \dots B_{i_N}^g (\la_N)
	   \ph_g^{i_1 \dots i_N} (\m_1, \dots, \m_M) \Om \epp
\end{equation}
Here the $B_j^g$, $j = 1, 2$, may be thought of as components of row
vectors
\begin{subequations}
\begin{align}
     B^{(+--)} (\la) & = \bigl(\CT^1_2 (\la), \CT^1_3 (\la)\bigr)
                           \epc \\
     B^{(-+-)} (\la) & = \bigl(\CT^1_3 (\la), \CT^1_2 (\la)\bigr)
                           \epc \\
     B^{(--+)} (\la) & = \bigl(\CT^1_2 (\la), \CT^1_3 (\la)\bigr) \epp
\end{align}
\end{subequations}
$\ph_g$ is defined in terms of an auxiliary monodromy matrix which
is a product of two $2 \times 2$-matrices
\begin{equation} \label{defttilde}
     \tilde \CT (\la) = \begin{pmatrix}
		        \tilde A (\la) & \tilde B (\la) \\
			\tilde C (\la) & \tilde D (\la)
			\end{pmatrix}
		      = {\cal D}_g (\la) \CT_{g'} (\la) \epp
\end{equation}
The factor ${\cal D}_g (\la)$ basically contains elements of the
monodromy matrix $\CT (\la)$,
\begin{subequations}
\label{dies}
\begin{align}
     {\cal D}_{(+--)} (\la) & = \begin{pmatrix}
                                \CT^2_2 (\la) & \CT^2_3 (\la) \\[1ex]
			        \CT^3_2 (\la) & \CT^3_3 (\la)
			        \end{pmatrix} \epc \\[1ex]
     {\cal D}_{(-+-)} (\la) & = \begin{pmatrix}
                                \CT^3_3 (\la) & \CT^3_2 (\la)
			        (\s^z)^{\otimes N} \\[1ex]
			        \CT^2_3 (\la) (\s^z)^{\otimes N}
				& \CT^2_2 (\la)
			        \end{pmatrix} \epc \label{dmpm} \\[1ex]
     {\cal D}_{(--+)} (\la) & = \begin{pmatrix}
                                \CT^2_2 (\la) & \CT^2_3 (\la)
			        (\s^z)^{\otimes N} \\[1ex]
			        \CT^3_2 (\la) (\s^z)^{\otimes N}
				& \CT^3_3 (\la)
			        \end{pmatrix} \epp \label{dmmp}
\end{align}
\end{subequations}
The factor $\CT_{g'} (\la)$ is the monodromy matrix of an auxiliary
`spin problem',
\begin{equation} \label{deftprime}
     \CT_{g'} (\la) = \CL_N^{g'} (\la_N - \la) \dots
                      \CL_1^{g'} (\la_1 - \la) \epc
\end{equation}
carrying an induced grading $g'$, which is $(+,+)$ for $g = (+,-,-)$
and $(+,-)$ for the remaining two cases $g = (-,+,-)$ and $g = (-,-,+)$.
The corresponding elementary $L$-matrices are\footnote{Note that, as
compared to \cite{Goehmann02}, we have changed the definition of the
$L$-matrix $\CL_j^{(++)}$ into the equivalent form (\ref{lpp}).}
\begin{equation} \label{lpp}
     \CL_j^{(++)} (\la) = a(\la) I_2 + b(\la) \begin{pmatrix}
                     {e_j}_1^1 & {e_j}_2^1 \\[1ex]
		     {e_j}_1^2 & {e_j}_2^2
		     \end{pmatrix} \epp
\end{equation}
and
\begin{equation} \label{lpm}
     \CL_j^{(+-)} (\la) = a(\la) I_2 + b(\la) \begin{pmatrix}
                     {e_j}_1^1 & {e_j}_2^1 \\[1ex]
		     {e_j}_1^2 &  - {e_j}_2^2
		     \end{pmatrix} \epp
\end{equation}
The ${e_j}_\a^\be$ are the canonical basis elements of $\bigl(
\End({\mathbb C}^2) \bigr)^{\otimes N}$ (viewed as a graded algebra)
introduced in appendix~\ref{app:defga}. They depend on the induced
grading $g'$. The $L$-matrix $\CL_j^{(++)}$ is the $L$-matrix of the
XXX spin chain of spin $\2$, whereas $\CL_j^{(+-)}$ generates the
gl(1$|$1) invariant model of free spinless fermions.

Inserting (\ref{dies})-(\ref{lpm}) into (\ref{defttilde}) we have
defined the auxiliary monodromy matrix $\tilde \CT (\la)$ for the
three different gradings under consideration. The $2^N$-column vector
$\ph_g$ in equation (\ref{evform}) is constructed by acting with
matrix elements of $\tilde \CT (\la)$ on appropriate auxiliary states.
For $g = (+,-,-)$ and $g = (-,-,+)$ we define
\begin{equation}
     \ph_g (\m_1, \dots, \m_M) = \tilde B (\m_1) \dots \tilde B (\m_M)
                                 \tst{\binom{1}{0}^{\otimes N}} \epc
\end{equation}
for $g = (-,+,-)$ an appropriate definition is
\begin{equation}
     \ph_g (\m_1, \dots, \m_M) = \tilde C (\m_1) \dots \tilde C (\m_M)
                                 \tst{\binom{0}{1}^{\otimes N}} \epp
\end{equation}
A derivation of the eigenvectors and the corresponding eigenvalues
for $g = (-,+,-)$ and $g = (-,-,+)$ is presented in appendix~%
\ref{app:derev}. The proof for $g = (+,-,-)$ can be found in
\cite{Goehmann02}.

\section{Examples} \label{sec:examples}
Numerous examples of systems for which the Bethe ansatz solution of
the previous section applies can be constructed starting from the
observation \cite{Kulish85} that
\begin{equation} \label{genl}
     {\CL_j}^\a_\be (\la) = a(\la) \de^\a_\be + b(\la)
                            (-1)^{p(\a) p(\be)} {E_j}^\a_\be
\end{equation}
with $a(\la), b(\la)$ defined in (\ref{defab}) is a representation of
the graded Yang-Baxter algebra (\ref{gyba}) if ${E_j}^\a_\be$ is a
representation of gl(1$|$2) of parity compatible with the grading,
\begin{equation} \label{gl12}
     [{E_j}^\a_\be, {E_k}^\g_\de]_\pm =
        \de_{jk} \bigl( \de^\a_\de {E_j}^\g_\be
	              - (-1)^{(p(\a) + p(\be))(p(\g) + p(\de))}
                        \de^\g_\be {E_j}^\a_\de \bigr) \epp
\end{equation}
Here $[\cdot, \cdot]_\pm$ denotes the superbracket (see
appendix~\ref{app:defga}).

For the construction of models of fermions on one-dimensional
lattices (which is our personal concern with the Bethe ansatz presented
in the previous chapter) one may generally utilize (\ref{genl}) in a
way that involves three logically separate steps:
\begin{enumerate}
\item
Take a representation of gl(1$|$2) in $\End( {\mathbb C}^n )$ (eq.\
(\ref{gl12}) with $j = k$, see \cite{SNR77,Marcu80}).
\item
Embed it into $\bigl( \End( {\mathbb C}^n) \bigl)^{\otimes L}$ in such
a way that the grading ((\ref{gl12}) for $j \ne k$) is respected
\cite{GoMu98}.
\item
Introduce Fermi operators \cite{GoMu98,GoKo00}.
\end{enumerate}
Since we merely want to illustrate our Bethe ansatz solution of the
previous sections we shall take the three steps in one in the examples
considered below.

The most elementary example is, of course, the supersymmetric $t$-$J$
model \cite{Schlottmann87,EsKo92} which is the fundamental model
associated with the $R$-matrix (\ref{rm}). The supersymmetric $t$-$J$
model is a model of electrons on a lattice. In order to be able to
write down the $L$-matrix and the Hamiltonian in a familiar way we
introduce canonically anticommuting creation and annihilation
operators $c_{j, a}^\+$, $c_{k, b}$ where the indices $j, k = 1,
\dots, L$ refer to the lattice sites, and $a, b = \auf, \ab$ are spin
indices.

Due to the canonical anticommutation relations the elements
$(X_j)^\a_\be$, $\a, \be = 1, 2, 3$, of the matrix
\begin{equation} \label{xjtj}
     X_j = \begin{pmatrix}
	     (1 - n_{j \ab})(1 - n_{j \auf}) &
	     (1 - n_{j \ab}) c_{j \auf} &
	     c_{j \ab} (1 - n_{j \auf}) \\[.5ex]
	     (1 - n_{j \ab}) c_{j \auf}^\+ &
	     (1 - n_{j \ab}) n_{j \auf} &
	     - c_{j \ab} c_{j \auf}^\+ \\[.5ex]
	     c_{j \ab}^\+ (1 - n_{j \auf}) &
	     c_{j \ab}^\+ c_{j \auf} &
	     n_{j \ab} (1 - n_{j \auf})
           \end{pmatrix}
\end{equation}
with $n_{j, \auf} = c_{j, \auf}^\+ c_{j, \auf}$, $n_{j, \ab} =
c_{j, \ab}^\+ c_{j, \ab}$ form a complete set of `projection
operators' on the space of states locally spanned by the basis vectors
$|0\>$, $c_{j \auf}^\+ |0\>$, $c_{j \ab}^\+ |0\>$. Double occupancy of
lattice sites is forbidden on this space. Setting ${X_j}_\a^\be =
(X_j)^\a_\be$, $\a, \be = 1, 2, 3$, we find
\begin{subequations}
\label{project}
\begin{align}
     {X_j}_\a^\be {X_j}_\g^\de & = \de^\be_\g {X_j}_\a^\de \epc \\
     {X_j}_\a^\be {X_k}_\g^\de & =
        (-1)^{(p(\a) + p(\be))(p(\g) + p(\de))}
	{X_k}_\g^\de {X_j}_\a^\be \epc \qd \text{for $j \ne k$} \epc
\end{align}
\end{subequations}
where $p(1) = 0$, $p(2) = p(3) = 1$. It follows that the operators
${X_j}_\a^\be$ satisfy equation (\ref{gl12}). The linear combination
${X_j}_\a^\a = 1 - n_{j \auf} n_{j \ab}$ projects the local space of
lattice electrons onto the space from which double occupancy is
excluded. The corresponding global projection operator is
\begin{equation} \label{notwo}
     P_0 = \prod_{j=1}^L (1 - n_{j \auf} n_{j \ab}) \epp
\end{equation}
It will be needed below.

We conclude with (\ref{genl}), (\ref{gl12}) and (\ref{project}) that
the $L$-matrix
\begin{equation} \label{ltj}
     \CL_j (\la) = a(\la) I_3 + b(\la) \begin{pmatrix}
		    {X_j}_1^1 & {X_j}_2^1 & {X_j}_3^1 \\[.5ex]
		    {X_j}_1^2 & - {X_j}_2^2 & - {X_j}_3^2 \\[.5ex]
		    {X_j}_1^3 & - {X_j}_2^3 & - {X_j}_3^3
                 \end{pmatrix} \epp
\end{equation}
is a representation of the graded Yang-Baxter algebra (\ref{gyba})
with grading $(+,-,-)$. This representation has been termed fundamental
graded representation in \cite{GoMu98}. The Fock vacuum $|0\>$
satisfying $c_{j, a} |0\> = 0$ for $j = 1, \dots, L$; $a, b =
\auf, \ab$ is clearly a possible highest vector for $\CL_j (\la)$,
\begin{equation} \label{ltjvac}
     \CL_j (\la) |0\> = \begin{pmatrix}
		    1 & b(\la) {X_j}_2^1 & b(\la) {X_j}_3^1 \\[.5ex]
		    0 & a(\la) & 0 \\[.5ex]
		    0 & 0 & a(\la)
                 \end{pmatrix} |0\> \epp
\end{equation}

It turns out that the matrix $\CL_j (\la)$ generates the supersymmetric
$t$-$J$ model at a single site. The corresponding monodromy matrix of
the $L$-site model is
\begin{equation} \label{ttj}
     \CT (\la) = \CL_L (\la) \dots \CL_1 (\la) \epp
\end{equation}
Its action on the Fock vacuum follows from (\ref{ltjvac}) as
\begin{equation} \label{ttjvac}
     \CT (\la) |0\> = \begin{pmatrix}
		    1 & B_1 (\la) & B_2 (\la) \\[.5ex]
		    0 & a^L (\la) & 0 \\[.5ex]
		    0 & 0 & a^L (\la)
                 \end{pmatrix} |0\> \epp
\end{equation}
Thus, $\CT (\la)$ is a representation of the graded Yang-Baxter
algebra (\ref{gyba}), and $|0\>$ is a highest vector satisfying
(\ref{para}). It follows that our general formulae (\ref{lapmm}),
(\ref{bapmm}) apply with functional parameters which can be read
off from (\ref{ttjvac}):
\begin{equation}
     a_1 (\la) = 1 \epc \qd a_2 (\la) = a_3 (\la) = a^L (\la) \epp
\end{equation}
This way we have recovered equations (3.47), (3.48) and (3.50) of
\cite{EsKo92}.

Note that the Hamiltonian of the supersymmetric $t$-$J$ model is
\begin{equation}
     H = - \i c \, \6_\la \ln \bigl\{ \bigl( \str (\CT (0)) \bigr)^{-1}
                                       \str (\CT (\la)) \bigr\}
				       \Bigr|_{\la = 0} \epp
\end{equation}
Because it acts on the restricted space of electronic states, where no
lattice site is doubly occupied, we may replace it with (see
\cite{GoMu98,GoKo00})
\begin{equation}
     H P_0 = P_0 \Bigl\{ - \sum_{j=1}^L
        (c_{j, a}^\+ c_{j+1, a}^{} + c_{j+1, a}^\+ c_{j, a}^{})
	+ 2 \sum_{j=1}^L \left( S_j^\a S_{j+1}^\a
	- \frac{n_j n_{j+1}}{4} + n_j \right) \Bigr\} P_0
	\epc
\end{equation}
where we inserted the usual definitions $S_j^\a =
\2 \sum_{a, b = \auf, \ab} c_{j, a}^\+ \s^\a_{ab} c_{j, b}^{}$
of local spin operators in terms of Pauli matrices $\s^\a$ and
introduced the local particle number operators $n_j = n_{j, \auf} +
n_{j, \ab}$.

Clearly the Fock vacuum is also a highest vector for $\Pi_{23} \CT (\la)
\Pi_{23}$, since under a permutation of the second and third basis
vector in auxiliary space $\CL_j (\la) |0\>$ transforms into
\begin{equation} \label{ltjvacp23}
     \Pi_{23} \CL_j (\la) \Pi_{23} |0\> =
                 \begin{pmatrix}
		    1 & b(\la) {X_j}_3^1 & b(\la) {X_j}_2^1 \\[.5ex]
		    0 & a(\la) & 0 \\[.5ex]
		    0 & 0 & a(\la)
                 \end{pmatrix} |0\> \epp
\end{equation}
which is of upper triangular form. The monodromy matrices
$\CT (\la)$ and $\Pi_{23} \CT (\la) \Pi_{23}$ carry the same grading
and, as for arbitrary permutations, lead to the same transfer matrix
(see equation (\ref{tperminv})). Since the grading as well as the
parameters $a_1$, $a_2$ and $a_3$ are identical for $\CT (\la)$ and
$\Pi_{23} \CT (\la) \Pi_{23}$, both monodromy matrices lead to the
same form (\ref{lapmm}) of the transfer matrix eigenvalue and to the
same Bethe ansatz equations (\ref{bapmm}). Note, however, that the
Bethe ansatz eigenvectors (\ref{evform}) are different, because
$B^{(+--)} (\la)$ is changed to $B^{(+--)} (\la) \s^x$ which is
equivalent to a spin flip transformation.

Indeed, applying the spin flip transformation
\begin{equation}
     J^{(s)} = \prod_{j=1}^L \bigl(1 - (c_{j, \auf}^\+ - c_{j, \ab}^\+)
                  (c_{j, \auf} - c_{j, \ab})\bigr)
\end{equation}
to the elementary $L$-matrix we obtain
\begin{equation}
     J^{(s)} \CL_j (\la) \bigl( J^{(s)} \bigr)^\+
        = \Pi_{23} \CL_j (\la) \Pi_{23}
\end{equation}
which implies the invariance of the transfer matrix with respect to
spin flips.

It is a well known fact that there are three alternative sets of
Bethe ansatz equations for the supersymmetric $t$-$J$ model
\cite{EsKo92,FoKa93}. Let us see how this comes out in our general
formalism. We shall consider the monodromy matrix action on the two
states
\begin{subequations}
\begin{align}
     |\auf \:\> = {X_L}_2^1 \dots {X_1}_2^1 |0\>
                = c_{L, \auf}^\+ \dots c_{1, \auf}^\+ |0\> \epc \\[1ex]
     |\ab \:\> = {X_L}_3^1 \dots {X_1}_3^1 |0\>
                = c_{L, \ab}^\+ \dots c_{1, \ab}^\+ |0\> \epp
\end{align}
\end{subequations}
Calculating first of all the action of the $L$-matrix on these states
we obtain
\begin{align}
     \CL_j (\la) |\auf \:\> & =
                 \begin{pmatrix}
		    a(\la) & 0 & 0 \\[.5ex]
		    b(\la) {X_j}_1^2 & a(\la) - b(\la) &
		    - b(\la) {X_j}_3^2 \\[.5ex]
		    0 & 0 & a(\la)
                 \end{pmatrix} |\auf \:\> \epc \\[1ex]
     \CL_j (\la) |\ab \:\> & =
                 \begin{pmatrix}
		    a(\la) & 0 & 0 \\[.5ex]
		    0 & a(\la) & 0 \\[.5ex]
		    b(\la) {X_j}_1^3 & - b(\la) {X_j}_2^3
		    & a(\la) - b(\la)
                 \end{pmatrix} |\ab \:\> \epp
\end{align}
\begin{figure}

\tabcolsep15pt
\renewcommand{\arraystretch}{1.8}

\begin{center}

\begin{tabular}{|c|c|c|c|c|c|} \hline
$\s \in {\mathfrak S}^3$ & $\Om$ & $g$ & $a_1 (\la)$ & $a_2 (\la)$ &
$a_3 (\la)$ \\ \hline \hline
$\id$ & $|0\>$ & $(+,-,-)$ & $1$ & $a^L (\la)$ & $a^L (\la)$ \\
\hline
$\Pi_{12}$ & $|\auf \:\>$ & $(-,+,-)$ & $\frac{a^L (\la)}{a^L (- \la)}$
& $a^L (\la)$ & $a^L (\la)$ \\ \hline
$\Pi_{12} \Pi_{23}$ & $|\ab \:\>$ & $(-,+,-)$ &
$\frac{a^L (\la)}{a^L (- \la)}$ & $a^L (\la)$ & $a^L (\la)$ \\ \hline
$\Pi_{12} \Pi_{23} \Pi_{12}$ & $|\ab \:\>$ & $(-,-,+)$ &
$\frac{a^L (\la)}{a^L (- \la)}$ & $a^L (\la)$ & $a^L (\la)$ \\ \hline
$\Pi_{23} \Pi_{12}$ & $|\auf \:\>$ & $(-,-,+)$ &
$\frac{a^L (\la)}{a^L (- \la)}$ & $a^L (\la)$ & $a^L (\la)$ \\ \hline
$\Pi_{23}$ & $|0\>$ & $(+,-,-)$ & $1$ & $a^L (\la)$ & $a^L (\la)$ \\
\hline
\end{tabular}

\end{center}

\caption{Different Bethe ansatz solutions for supersymmetric $t$-$J$
model.}
\label{fig:tab}
\end{figure}
Comparing the patterns of zeros on the right hand side of these
equations to figure \ref{fig:permon} we see that $|\auf \:\>$ is
a highest vector for $\Pi_{12} \CT (\la) \Pi_{12}$ and for
$\Pi_{23} \Pi_{12} \CT (\la) \Pi_{12} \Pi_{23}$, while $|\ab \:\>$
is a highest vector for $\Pi_{12} \Pi_{23} \CT (\la) \Pi_{23}
\Pi_{12}$ and for $\Pi_{12} \Pi_{23} \Pi_{12} \CT (\la) \Pi_{12}
\Pi_{23} \Pi_{12}$. Together with the possibilities already covered
by using $|0\>$ as a highest vector we obtain all six cases of
figure \ref{fig:permon} albeit with different grading. The situation
is summarized in tabular~\ref{fig:tab}.

Taking the data from the tabular and inserting it into (\ref{laggg})-%
(\ref{evform}) we obtain the different possible Bethe ansatz solutions
of the supersymmetric $t$-$J$ model. The expressions for the eigenvalues
and Bethe ansatz equations are in agreement with the results of
\cite{EsKo92}. Because of space limitations we do not repeat those
results here. In \cite{EsKo92} and also in \cite{FoKa93} the authors
avoided writing explicit expressions for the $L$-matrix in terms of
Fermi operators. Therefore they could not see the correspondence
between the possible choices of pseudo vacua $|0\>$, $|\auf \:\>$,
$|\ab \:\>$ and the possible gradings. As we can learn from tabular~%
\ref{fig:tab} this correspondence is not unique. For the highest vector
$|\auf \:\>$ the Bethe ansatz can be realized with grading $(-,+,-)$
or $(-,-,+)$, respectively. A similar statement holds for $|\ab \:\>$.
By way of contrast, the Bethe ansatz equations and the expression for
the transfer matrix eigenvalue are uniquely fixed if we choose $|0\>$
as the highest vector. Still, as was observed above, the eigenvectors
can be realized in two different ways.

The algebraic Bethe ansatz for the supersymmetric $t$-$J$ model is
rather special as compared to the general case. This is due to the
fact that $a_2 (\la) = a_3 (\la)$ (see tabular \ref{fig:tab}) and
that an additional zero appears above the diagonal in the action of
the monodromy matrix on the highest vector. Consider, for instance,
the case $\Om = |0\>$, $\s = \id$. Then the monodromy matrix action on
$\Om$ is given by (\ref{ttjvac}). Making contact with the notation
of (\ref{dies}) we see that ${\cal D}_{(+--)} (\la) |0\> = a^L (\la)
I_2 |0\>$. Because of this trivial action we may drop the factor
${\cal D}_g (\la)$ on the right hand side of (\ref{defttilde}) and the
eigenvectors (\ref{evform}) are written only in terms of
$\CT^1_2 (\la)$, $\CT^1_3 (\la)$ and the auxiliary monodromy matrix
$\CT_{(++)} (\la)$. (see (\ref{deftprime}), (\ref{lpp})).

More examples are obtained by inserting other representations of
(\ref{gl12}) into (\ref{genl}). For the grading $g = (+,-,-)$ we may,
for instance, take the four-dimensional representation
\begin{equation}
     E = \begin{pmatrix}
            \ch^2 (\a) - n_\auf - n_\ab &
	    \ch (\a) \, c_\auf^\+ - e^{- \a} c_\auf^\+ n_\ab &
	    \ch (\a) \, c_\ab^\+ - e^{- \a} c_\ab^\+ n_\auf \\
	    \ch (\a) \, c_\auf - e^{- \a} c_\auf n_\ab &
	    n_\auf & c_\ab^\+ c_\auf \\
	    \ch (\a) \, c_\ab - e^{- \a} c_\ab n_\auf &
	    c_\auf^\+ c_\ab & n_\ab
	 \end{pmatrix}
\end{equation}
which depends on a free parameter $\a \in {\mathbb R}$. Note however,
that it requires more effort \cite{KuSk82b} to associate a physically
relevant model having a local Hamiltonian with higher dimensional
representations. This is a subject that exceeds the scope of this work.

\section{Conclusion}
We have completed the task, begun in \cite{Goehmann02}, of constructing
the algebraic Bethe ansatz for the gl(1$|$2) generalized model.
In this work the gradings $g = (-,+,-)$ and $g = (-,-,+)$ have been
treated. As we hope to have convinced the reader in sections 3 and 5,
a complete understanding requires to consider the three gradings
together.

We hope our work will prove to be useful in future constructions
of exact solutions of models with $R$-matrix (\ref{rm}). In first place
we think of novel impurity models and of possible applications to
Yang's model of electrons interacting via delta function potential and,
maybe, to the Hubbard model (see discussion in \cite{Goehmann02}).
Other applications may be the calculation of quantum transfer matrix
eigenvalues (the quantum transfer matrix has a staggered pseudo vacuum
on which the monodromy matrix acts `without producing additional zeros
above the diagonal') and the calculation of norms of Bethe ansatz
eigenstates (see \cite{Goehmann02}).

\newpage

{\appendix
\section{Graded algebras} \label{app:defga}

In this appendix we shall recall the basic concepts of graded vector
spaces and graded associative algebras. In the context of the quantum
inverse scattering method these concepts were first utilized by Kulish
and Sklyanin \cite{KuSk80,Kulish85}.

Graded vector spaces are vector spaces equipped with a notion of odd
and even, that allows us to treat fermions within the formalism of the
quantum inverse scattering method (see \cite{GoMu98,GoKo00}). Let us
consider a finite dimensional vector space $V$, which is the direct
sum of two subspaces, $V = V_0 \oplus V_1$, $\dim V_0 = m$,
$\dim V_1 = n$. We shall call $v_0 \in V_0$ even and $v_1 \in V_1$
odd. The subspaces $V_0$ and $V_1$ are called the homogeneous components
of $V$. The parity $\p$ is a function $V_i \rightarrow \mathbb{Z}_2$
defined on the homogeneous components of~$V$,
\begin{equation}
     \p(v_i) = i \epc \qd i = 0, 1 \epc \qd v_i \in V_i \epp
\end{equation}
The vector space $V$ endowed with this structure is called a graded
vector space or super space. 

Let ${\cal A}$ be an associative algebra (with unity), which is graded
as a vector space. Suppose $X, Y \in {\cal A}$ are homogeneous. If the
product $XY$ is homogeneous with parity
\begin{equation} \label{homab}
     \p(XY) = \p(X) + \p(Y) \epc
\end{equation}
then ${\cal A}$ is called a graded associative algebra \cite{KuSk80}.

For any two homogeneous elements $X, Y \in {\cal A}$ let us define the
super-bracket
\begin{equation} \label{superbracket}
     [X,Y]_\pm = XY - (-1)^{\p(X)\p(Y)} YX \epc
\end{equation}
and let us extend this definition linearly in both of its arguments
to all elements of~${\cal A}$.

Let $p: \{1, \dots, n \} \rightarrow {\mathbb Z}_2$. The set of all
$n \times n$ matrices $A, B, C, \dots$ with entries in ${\cal A}$, such
that $\p (A^\a_\be) = \p (B^\a_\be) = \p (C^\a_\be) = \dots = p(\a) +
p(\be)$ is an associative algebra, say ${\rm Mat} ({\cal A}, n)$, since
$\p (A^\a_\be B^\be_\g) = p(\a) + p(\g)$. For $A, B \in {\rm Mat}
({\cal A}, n)$ we define the graded tensor product (or super tensor
product)
\begin{equation} \label{defgtp}
     (A \otimes_g B)^{\a \g}_{\be \de} = (-1)^{(p(\a) + p(\be)) p(\g)}
                                            A^\a_\be B^\g_\de \epp
\end{equation}
The graded tensor product is associative. For matrices $A, B, C, D \in
{\rm Mat} ({\cal A}, n)$ with $[B^\a_\be, C^\g_\de]_\pm = 0$ we find
\begin{equation} \label{abcd}
     (A \otimes_g B)(C \otimes_g D) = AC \otimes_g BD \epp
\end{equation}

Our chief example of a graded associative algebra is the algebra
${\cal A} = (\End(V))^{\otimes L}$, where $V$ is a graded vector space
as introduced above. ${\cal A}$ acquires the structure of a graded
algebra in the following way: We fix a basis $\{e_1, \dots, e_{m+n}\}$
of definite parity and define $p: \{1, \dots m + n\} \rightarrow
{\mathbb Z}_2$ by setting $p(\a) = \pi(e_\a)$. Then the set $\{e_\a^\be
\in \End(V) | \a, \be = 1, \dots m + n\}$ with $e_\a^\be e_\g =
\de_\g^\be e_\a$ is a basis of $\End(V)$, and the tensor products
$e_{\a_1}^{\be_1} \otimes \dots \otimes e_{\a_L}^{\be_L}$ span the
vector space ${\cal A} = (\End(V))^{\otimes L}$. Hence, the definition
\begin{equation} \label{gradendl}
     \p (e_{\a_1}^{\be_1} \otimes \dots \otimes e_{\a_L}^{\be_L})
               = p(\a_1) + p(\be_1) + \dots + p(\a_L) + p(\be_L)
\end{equation}
induces a grading on ${\cal A}$ regarded as a vector space. It is
easy to see that an element $X = X^{\a_1 \dots \a_L}_{\be_1 \dots \be_L}
e_{\a_1}^{\be_1} \otimes \dots \otimes e_{\a_L}^{\be_L} \in {\cal A}$
is homogeneous with parity $\pi (X)$ if and only if
\begin{equation}
     (-1)^{\sum_{j=1}^L (p(\a_j) + p(\be_j))}
        X^{\a_1 \dots \a_L}_{\be_1 \dots \be_L} =
	   (-1)^{\p(X)} X^{\a_1 \dots \a_L}_{\be_1 \dots \be_L} \epp
\end{equation}
But the latter equation implies that two homogenous elements $X$ and
$Y$ satisfy equation (\ref{homab}), and ${\cal A}$ is a graded algebra.

The following definition of `graded local projection operators'
\cite{GoMu98} provides a standard basis of the graded associative
algebra ${\cal A}$ which is of crucial importance in constructing
solvable lattice models,
\begin{equation} \label{defej}
     {e_j}_\a^\be = (-1)^{(p(\a) + p(\be)) \sum_{k= 1}^{j-1} p(\g_k)}
		    \, e_{\g_{1}}^{\g_{1}} \otimes \dots
		    \otimes e_{\g_{j-1}}^{\g_{j-1}}
		    \otimes e_\a^\be \otimes I_{m + n}^{\otimes (L-j)}
		    \epp
\end{equation}
Here $I_{m + n}$ is the $(m + n) \times (m + n)$ unit matrix, and
summation over double tensor indices (i.e., over $\g_{1}, \dots,
\g_{j-1}$) is implied. The index $j$ on the left hand side of
(\ref{defej}) is called the site index. A simple consequence of the
definition (\ref{defej}) for $j \ne k$ are the commutation relations
\begin{equation} \label{coantico}
     {e_j}_\a^\be {e_k}_\g^\de = (-1)^{(p(\a) + p(\be))(p(\g) + p(\de))}
          {e_k}_\g^\de {e_j}_\a^\be \epp
\end{equation}
It further follows from equation (\ref{defej}) that ${e_j}_\a^\be$ is
homogeneous with parity
\begin{equation}
     \p({e_j}_\a^\be) = p(\a) + p(\be) \epp
\end{equation}
Hence, equation (\ref{coantico}) says that odd matrices with different
site indices mutually anticommute, whereas even matrices commute with
each other as well as with the odd matrices. For products of matrices
${e_j}_\a^\be$ which are acting on the same site (\ref{defej}) implies
the projection property
\begin{equation} \label{samesite}
     {e_j}_\a^\be {e_j}_\g^\de = \de_\g^\be {e_j}_\a^\de \epp
\end{equation}
Using the super-bracket (\ref{superbracket}), equations
(\ref{coantico}) and (\ref{samesite}) can be combined into
\begin{equation} \label{imbglmn}
     [{e_j}_\a^\be,{e_k}_\g^\de]_\pm =
	  \de_{jk} \left( \de_\g^\be {e_j}_\a^\de -
	  (-1)^{(p(\a) + p(\be))(p(\g) + p(\de))}
          \de_\a^\de {e_j}_\g^\be \right) \epp
\end{equation} 
The right hand side of the latter equation with $j = k$ gives the
structure constants of the Lie super algebra gl($m|n$) with respect
to the basis $\{{e_j}_\a^\be\}$.

Since any $m + n$-dimensional vector space over the complex numbers
is isomorphic to $\mathbb{C}^{m+n}$, we may simply set $V =
\mathbb{C}^{m+n}$. We may further assume that our homogeneous basis
$\{ e_\a \in \mathbb{C}^{m+n}| \a = 1, \dots, m + n \}$ is canonical,
i.e., we may represent the vector $e_\a$ by a column vector having the
only non-zero entry +1 in row $\a$. Our basic matrices $e_\a^\be$ are
then $(m + n) \times (m + n)$-matrices with a single non-zero entry +1
in row $\a$ and column $\be$.

The definition (\ref{defej}) generalizes the notion of the Jordan-Wigner
transformation to systems with higher spin (see \cite{GoKo00}). As
with the Jordan-Wigner transformation another consistent definition
of the graded local projection operators, also leading to
(\ref{coantico}) and (\ref{samesite}), is obtained by placing the
factors $(-1)^{(p(\a) + p(\be))p(\g_k)} e_{\g_k}^{\g_k}$ in the tensor
product on the right hand side of (\ref{defej}) behind rather than in
front of $e_\a^\be$. This alternative convention was used in
\cite{GoMu98,GoKo00}. Here we use (\ref{defej}) instead as it
naturally appears in the derivation of the algebraic Bethe ansatz
for the gl(1$|$2) generalized model with gradings $g = (-,+,-)$ and
$g = (-,-,+)$ (see appendix~\ref{app:derev}).

\section{Derivation of the eigenvectors and eigenvalues of the transfer
matrix for ${\mathbf g = (-,+,-)}$ and ${\mathbf g = (-,-,+)}$}
\label{app:derev}
It is most convenient to start with the case $g = (-,-,+)$, for which
the Yang-Baxter algebra has a simple block structure, and to obtain
the case $g = (-,+,-)$ afterwards. In fact, it is equivalent to the
case $g = (-,-,+)$ but with a transfer matrix acting on the pseudo
vacuum $\Om$ as
\begin{equation} \label{vacactmpm}
     \CT (\la) \Om = \begin{pmatrix}
                        a_1 (\la) & \ast & \ast \\
			0 & a_3 (\la) & 0 \\
			0 & \ast & a_2 (\la)
		     \end{pmatrix} \Om \epp
\end{equation}
The `first level algebraic Bethe ansatz' will be the same in both cases.
The difference comes in only on the second level.

The first step of our calculation is to rewrite the graded Yang-Baxter
algebra (\ref{gyba}) with $R$-matrix $\check R_{(--+)} (\la)$ in block
form: We introduce the shorthand notations
\begin{equation}
\begin{split}
     & B(\la) = \big( B_1 (\la), B_2 (\la) \big) \epc \qd
       C(\la) = \begin{pmatrix} C^1 (\la) \\ C^2 (\la) \end{pmatrix}
                   \epc \\
     & D(\la) = \begin{pmatrix} D^1_1 (\la) & D^1_2 (\la) \\
                                D^2_1 (\la) & D^2_2 (\la) \end{pmatrix}
				\epp
\end{split}
\end{equation}
Then the $3 \times 3$ monodromy matrix $\CT (\la)$ can be written as
\begin{equation}
     \CT (\la) = \begin{pmatrix} A (\la) & B (\la) \\
		                 C (\la) & D (\la)
				 \end{pmatrix} =
                 \begin{pmatrix} A (\la) & B_1 (\la) & B_2 (\la) \\
                                 C^1 (\la) & D^1_1 (\la) & D^1_2 (\la)
				 \\
				 C^2 (\la) & D^2_1 (\la) & D^2_2 (\la)
				 \end{pmatrix} \epp
\end{equation}
The defining relations of the graded Yang-Baxter algebra (\ref{gyba})
can be read as a $9 \times 9$ matrix equation. Let us denote the
$n \times n$ unit matrix by $I_n$. A similarity transformation with the
matrix
\begin{equation}
     X = \begin{pmatrix} I_4 & & \\
            & \begin{pmatrix} 0 & 0 & 1 \\ 1 & 0 & 0 \\ 0 & 1 & 0
	      \end{pmatrix} & \\
	    & & I_2
         \end{pmatrix} \epc
\end{equation}
which cyclically permutes the 5th, 6th and 7th row and column, followed
by a similarity transformations which multiplies the 5th, 8th and 9th
row and column by $-1$, transforms this $9 \times 9$ equation into
\begin{multline} \label{blockyba}
     \begin{pmatrix}
        1 & & & \\
	& b \, I_2 & a \, I_2 & \\
	& a \, I_2 & b \, I_2 & \\
	& & & \check r
     \end{pmatrix}
     \begin{pmatrix}
        A \otimes_{g'} \bar A & A \otimes_{g'} \bar B &
        B \otimes_{g'} \bar A & B \otimes_{g'} \bar B \\
        A \otimes_{g'} \bar C & A \otimes_{g'} \bar D &
        B \otimes_{g'} \bar C & B \otimes_{g'} \bar D \\
        C \otimes_{g'} \bar A & C \otimes_{g'} \bar B &
        D \otimes_{g'} \bar A & D \otimes_{g'} \bar B \\
        C \otimes_{g'} \bar C & C \otimes_{g'} \bar D &
        D \otimes_{g'} \bar C & D \otimes_{g'} \bar D
     \end{pmatrix} = \\[1ex]
     \begin{pmatrix}
        \bar A \otimes_{g'} A & \bar A \otimes_{g'} B &
        \bar B \otimes_{g'} A & \bar B \otimes_{g'} B \\
        \bar A \otimes_{g'} C & \bar A \otimes_{g'} D &
        \bar B \otimes_{g'} C & \bar B \otimes_{g'} D \\
        \bar C \otimes_{g'} A & \bar C \otimes_{g'} B &
        \bar D \otimes_{g'} A & \bar D \otimes_{g'} B \\
        \bar C \otimes_{g'} C & \bar C \otimes_{g'} D &
        \bar D \otimes_{g'} C & \bar D \otimes_{g'} D
     \end{pmatrix}
     \begin{pmatrix}
        1 & & & \\
	& b \, I_2 & a \, I_2 & \\
	& a \, I_2 & b \, I_2 & \\
	& & & \check r
     \end{pmatrix} \epp
\end{multline}
For the formula to fit on the line we suppressed the arguments and
adopted the following convention: $X = X (\la)$, $\bar X = X (\m)$ for
$X = A, \dots, D$. Moreover, $a = a(\m - \la)$ and $b = b(\m - \la)$.
The $4 \times 4$ matrix
\begin{equation}
     \check r = \begin{pmatrix} 1 & & & \\ & b & a & \\
                                & a & b & \\ & & & b - a
                \end{pmatrix}
\end{equation}
satisfies the Yang-Baxter equation. It is related to a special case
(rank 2, grading $(+,-)$) of the $R$-matrix (\ref{rm}) by equation
(\ref{defrcheck}) and is therefore unitary,
\begin{equation}
     \check r (\la) \check r (- \la) = I_4 \epp
\end{equation}
The grading $(+,-)$, corresponding to $p(1) = 0$, $p(2) = 1$, also
appears in the graded tensor products in (\ref{blockyba}), where it is
abbreviated as $g'$. These graded tensor products are defined by
equation (\ref{defgtp}) which makes not only sense for square matrices,
but for arbitrary $n \times m$ matrices. Thus, thinking of $B (\la)$
as a $1 \times 2$ matrix with row index $1$, and similarly of
$C (\la)$ as a $2 \times 1$ matrix with column index 1 and of $A (\la)$
as a $1 \times 1$ matrix with row and column index 1, all graded tensor
products in (\ref{blockyba}) are well defined. We have, for instance,
\begin{equation}
     B (\la) \otimes_{g'} C (\m) = \begin{pmatrix}
                                   B_1 (\la) C^1 (\m) &
				   B_2 (\la) C^1 (\m) \\
                                   B_1 (\la) C^2 (\m) &
				   - B_2 (\la) C^2 (\m)
				   \end{pmatrix} \epp
\end{equation}
We would like to remark that the defining relations of the graded
Yang-Baxter algebra of the gl(1$|$2) model, when written in block form
(\ref{blockyba}), resemble the corresponding relations for the gl(2)
model.

Out of the 16 relations contained in (\ref{blockyba}) we shall need the
following 4 for the first level algebraic Bethe ansatz,
\begin{align} \label{zamo}
     & B(\la) \otimes_{g'} B(\m) = \big( B(\m) \otimes_{g'} B(\la) \big)
                                \check r (\m - \la) \epc
				\\[1ex] \label{ab}
     & A(\la) \otimes_{g'} B(\m) = \frac{B(\m) \otimes_{g'} A(\la)}
                                     {a(\la - \m)}
				- \frac{b(\la - \m)}{a(\la - \m)} \,
				  B(\la) \otimes_{g'} A(\m)
			        \epc \\[1ex] \label{db}
     & D(\la) \otimes_{g'} B(\m) = \big( B(\m) \otimes_{g'} D(\la) \big)
                                \, \frac{\check r (\m - \la)}
				        {a(\m - \la)}
                                - \frac{b(\m - \la)}{a(\m - \la)} \,
				  B(\la) \otimes_{g'} D(\m)
				\epc \\[1ex] \label{dyba}
     & \check r (\m - \la) \big( D(\la) \otimes_{g'} D(\m) \big) =
       \big( D(\m) \otimes_{g'} D(\la) \big) \check r (\m - \la) \epp
\end{align}
Note that, by (\ref{zamo}), $B(\la)$ constitutes a representation of
the Zamolodchikov algebra, and, by (\ref{dyba}), $D(\la)$ is a
representation of the Yang-Baxter algebra of the gl(1$|$1) model.

Our goal is to calculate the eigenvectors of the transfer matrix
$t(\la) = - A(\la) - \str_{g'} (D(\la))$. In analogy with the gl(2)
case we shall first of all consider the commutation relations of a
multiple tensor product $B(\la_1) \otimes_{g'} \dots \otimes_{g'}
B(\la_N)$ with $A(\la)$ and $\str_{g'} (D(\la))$. These commutation
relations can be obtained by iterating equations (\ref{ab}) and
(\ref{db}):
\begin{align} \label{abmult}
     & A(\la) \Bigl[ \sideset{}{_{g'}} \bigotimes_{j=1}^N B(\la_j)
              \Bigr] = 
          \Bigl[ \sideset{}{_{g'}} \bigotimes_{j=1}^N B(\la_j) \Bigr]
	  A(\la) \prod_{j=1}^N \frac{1}{a(\la - \la_j)} \notag \\
        & \qqd - \sum_{j=1}^N \biggl\{ B(\la) \otimes_{g'}
	         \Bigl[
	         \sideset{}{_{g'}}
		 \bigotimes_{\substack{k=1 \\ k \ne j}}^N B(\la_k)
		 \Bigr] \biggr\} S_{j-1} \,
		 A(\la_j) \, \frac{b(\la - \la_j)}{a(\la - \la_j)}
		 \prod_{\substack{k=1 \\ k \ne j}}^N
		 \frac{1}{a(\la_j - \la_k)} \epc \\[1ex] \label{dbmult}
     & D(\la) \otimes_{g'}
           \Bigl[
	      \sideset{}{_{g'}} \bigotimes_{j=1}^N B(\la_j)
	   \Bigr] = \biggl\{
	      I_2 \otimes_{g'} \Bigl[ \sideset{}{_{g'}}
	           \bigotimes_{j=1}^N B(\la_j) \Bigr]
	      \biggr\}
	   \tilde \CT (\la) \prod_{j=1}^N \frac{1}{a(\la_j - \la)}
	       \notag \\
        & \qqd - \sum_{j=1}^N \biggl\{ I_2 \otimes_{g'}
	        B(\la) \otimes_{g'}
	      \Bigl[ \sideset{}{_{g'}}
	      \bigotimes_{\substack{k=1 \\ k \ne j}}^N
              B(\la_k) \Bigr] \biggr\} P_{01} (I_2 \otimes_{g'} S_{j-1})
              \cdot \notag \\[-1ex]
        & \mspace{180.0mu} \cdot \bigl\{ I_2 \otimes_{g'}
	        \str_{g'} \bigl(\tilde \CT (\la_j)\bigr) \bigr\}
		\, \frac{b(\la_j - \la)}{a(\la_j - \la)}
		\prod_{\substack{k=1 \\ k \ne j}}^N
		\frac{1}{a(\la_k - \la_j)} \epp
\end{align}
Here the operators $B(\la)$ in the multiple tensor products are
multiplied in ascending order. $\tilde \CT (\la)$ is defined in
equation (\ref{defttilde}). The operators $S_{j-1}$ appearing on the
right hand side of (\ref{abmult}) are given as
\begin{equation} \label{defsm}
     S_{j-1} = \bigl( \check r (\la_j - \la_1) \otimes_{g'}
                  I_2^{\otimes (N-2)} \bigr) \dots
               \bigl( I_2^{\otimes (j-2)} \otimes_{g'}
	          \check r (\la_j - \la_{j-1})
	          \otimes_{g'} I_2^{\otimes (N-j)} \bigr) \epc
\end{equation}
for $j = 2, \dots, N$. We further define $S_0 = \id$. The use of
the graded tensor product in (\ref{defsm}) makes sense, since all
non-zero matrix elements appear in such a way that they can be
interpreted as even elements of $\End ({\cal H})$. $P_{01}$ is a
graded transposition operator \cite{GoMu98} on $\bigl( \End
({\mathbb C}^2) \bigr)^{\otimes (N+1)}$ regarded as a graded algebra
(see appendix~\ref{app:defga}). In the canonical basis
$\{{e_0}_{\a_0}^{\be_0}, \dots, {e_N}_{\a_N}^{\be_N}\}$ it is
expressed as
\begin{equation}
     P_{jk} = (-1)^{p(\be)} {e_j}_\a^\be {e_k}_\be^\a \epc
\end{equation}
with $j = 0$ and $k = 1$. In general the operators $P_{jk}$ induce
the action of the symmetric group on the site indices of the basis
elements ${e_l}_\a^\be$.

Equations (\ref{abmult}) and (\ref{dbmult}) can be proven by induction
over $N$ (see \cite{Goehmann02}). As compared to the case $g =
(+,-,-)$ treated in \cite{Goehmann02} the main difference and the
main difficulty was to properly treat the graded tensor products
with the induced grading $g'$ in the derivation of equation
(\ref{dbmult}). The remaining part of the derivation is now similar to
the corresponding calculations for the case $g = (+,-,-)$. We shall
discuss it only briefly.

We take the super trace in space zero of equation (\ref{dbmult}) and
subtract it from minus one times equation (\ref{abmult}). Taking
into account that $t (\la) = - A (\la) - \str_{g'} (D (\la))$ for
the grading $g = (-,-,+)$ under consideration we obtain
\begin{align} \label{tbmult}
     & t(\la) \Bigl[ \sideset{}{_{g'}} \bigotimes_{j=1}^N B(\la_j)
              \Bigr] =
        \Bigl[ \sideset{}{_{g'}} \bigotimes_{j=1}^N B(\la_j) \Bigr]
	\notag \\[-1ex] & \mspace{168.0mu} \cdot
	  \Bigl\{
	     - A(\la) \prod_{j=1}^N \frac{1}{a(\la - \la_j)} -
	     \str_{g'} \bigr( \tilde \CT (\la) \bigl)
	     \prod_{j=1}^N \frac{1}{a(\la_j - \la)}
	  \Bigr\} \notag \\[1ex]
        & \qd + \sum_{j=1}^N \biggl( B(\la) \otimes_{g'}
	        \Bigl[ \sideset{}{_{g'}}
		\bigotimes_{\substack{k=1 \\ k \ne j}}^N B(\la_k) \Bigr]
		\biggr) S_{j-1} \,
		\frac{b(\la - \la_j)}{a(\la - \la_j)}
		\prod_{\substack{k=1 \\ k \ne j}}^N
		\frac{1}{a(\la_k - \la_j)}
		\notag \\[-1ex]
        & \mspace{168.0mu} \cdot
		\Bigl\{
		   A(\la_j) \prod_{\substack{k=1 \\ k \ne j}}^N
		   \frac{a(\la_k - \la_j)}{a(\la_j - \la_k)} -
		   \str_{g'} \bigl(\tilde \CT (\la_j)\bigr)
		\Bigr\} \epp
\end{align}
This is a form rather typical of an algebraic Bethe ansatz calculation
with wanted and unwanted terms on the right hand side of the equation.

The operator $\str_{g'} (\tilde \CT (\la))$ acts on the tensor
product $\bigl( {\mathbb C}^2 \bigr)^{\otimes N} \otimes {\cal H}$.
Its eigenvectors are independent of $\la$ since $\tilde \CT (\la)$
is a representation of the Yang-Baxter algebra with $R$-matrix
$\check r (\la)$,
\begin{equation} \label{gyba11}
     \check r (\m - \la)
        \big( \tilde \CT (\la) \otimes_{g'} \tilde \CT (\m) \big) =
	\big( \tilde \CT (\m) \otimes_{g'} \tilde \CT (\la) \big)
	\check r (\m - \la) \epp
\end{equation}
This holds first of all for $\CT_{g'} (\la)$ by construction
(see (\ref{deftprime})) and for $D (\la)$ by equation (\ref{dyba}).
It still holds after inserting the factors $(\s^z)^{\otimes N}$
into $D (\la)$ which then becomes ${\cal D} (\la)$ (see equation
(\ref{dies})). But, due to the factors $(\s^z)^{\otimes N}$ the
entries of ${\cal D} (\la)$ and $\CT_{g'} (\la)$ super-commute,
and (\ref{gyba11}) holds because of (\ref{abcd}).

Following Kulish and Reshetikhin \cite{KuRe83} we define the `vacuum
subspace' $\CH_0 \subset \CH$ by the conditions
\begin{subequations}
\begin{align} \label{h01}
     & A(\la) \PH = a_1 (\la) \PH \epc \\[.5ex]
     & C(\la) \PH = 0 \epc \label{h02}
\end{align}
\end{subequations}
for all $\PH \in \CH_0$. Clearly, $\CH_0$ is a linear subspace of $\CH$.
The following lemma \cite{Reshetikhin86} can be proven in a similar
manner as in \cite{Goehmann02}.
\begin{lemma}
$\CH_0$ is invariant under the action of $D(\la)$.
\end{lemma}
Equivalently we may say that the space spanned by all linear
combinations of vectors of the form $D^1_2 (\m_1) \dots D^1_2 (\m_M)
\, \Om$ is a linear subspace of $\CH_0$.

Suppose $\ph \, \Om \in \bigl( {\mathbb C}^2 \bigr)^{\otimes N} \otimes
{\cal H}_0$ is an eigenvector\footnote{We loosely write $\ph \, \Om$
instead of $\ph \otimes \Om$.} of $\str_{g'} (\tilde \CT (\la))$
with eigenvalue $\tilde \La (\la)$. Then $\ph \, \Om$ is a $2^N$-column
vector with entries in ${\cal H}_0$, and (\ref{h01}) holds for this
vector,
\begin{equation}
     A(\la) \ph \, \Om = a_1 (\la) \ph \, \Om \epp
\end{equation}
Thus, $\ph \, \Om$ is an eigenvector of the operators in curly
brackets on the right hand side of (\ref{tbmult}). Since the graded
tensor products of vectors $B (\la_j)$ form a $2^N$-row vector we
conclude
that
\begin{equation} \label{evpre}
     \Bigl[ \sideset{}{_{g'}} \bigotimes_{j=1}^N B(\la_j) \Bigr]
        \ph \, \Om = \sum_{i_1, \dots, i_N = 1}^2
	             B_{i_1} (\la_1) \dots B_{i_N} (\la_N)
		     \ph^{i_1, \dots, i_N} \, \Om \in {\cal H}
\end{equation}
is an eigenvector of the transfer matrix $t(\la)$ if the Bethe ansatz
equations
\begin{equation} \label{balpre}
     a_1 (\la_j) \prod_{\substack{k=1 \\ k \ne j}}^N
        \frac{a(\la_k - \la_j)}{a(\la_j - \la_k)} = \tilde \La (\la_j)
\end{equation}
are satisfied, which is just the condition for the unwanted terms in
the second curly bracket on the right hand side of (\ref{tbmult})
to vanish. The corresponding eigenvalue of $t(\la)$ is
\begin{equation} \label{tevpre}
     \La(\la) = - a_1 (\la) \prod_{j=1}^N \frac{1}{a(\la - \la_j)} -
                 \tilde \La (\la)
		 \prod_{j=1}^N \frac{1}{a(\la_j - \la)} \epp
\end{equation}

The remaining task is to solve the eigenvalue problem of $\str_{g'}
(\tilde \CT (\la))$ on the space $\bigl( {\mathbb C}^2
\bigr)^{\otimes N} \otimes {\cal H}_0$. This task can be accomplished
by a second Bethe ansatz which is possible, because $\tilde \CT (\la)$
is a representation of the Yang-Baxter algebra (see (\ref{gyba11})) and
the vector
\begin{equation} \label{vacmmp}
     \hat \Om = \tst{\binom{1}{0}^{\otimes N}} \Om
\end{equation}
is a highest vector for $\tilde \CT (\la)$. In fact, introducing the
explicit form of the $L$-matrix (\ref{lpm}) and the explicit form of
${\cal D_{(--+)}}$, equation (\ref{dmmp}), into the definition
(\ref{defttilde}) of $\tilde \CT (\la)$ we obtain
\begin{equation}
     \tilde \CT (\la) \, \hat \Om =
        \begin{pmatrix}
	   a_2 (\la) & \ast \\
	   0 & a_3 (\la) \prod_{j=1}^N a(\la_j - \la)
        \end{pmatrix} \hat \Om \epp
\end{equation}

For the construction of the eigenvectors of $\str_{g'}
(\tilde \CT (\la))$ we extract the following commutation relations from
(\ref{gyba11}),
\begin{subequations}
\label{tildecom}
\begin{align} \label{btbt}
     \tilde B (\la) \tilde B (\m) & = 
        \tilde B (\m) \tilde B (\la)
	\bigl(b(\m - \la) - a(\m - \la)\bigr) \epc \\[1ex] \label{adb}
     \bigl(\tilde A (\la) - \tilde D (\la)\bigr) \tilde B (\m) & = 
        \frac{\tilde B (\m) \bigl(\tilde A (\la) - \tilde D (\la)\bigr)}
	     {a(\la - \m)}
	- \frac{b(\la - \m)}{a(\la - \m)}
	   \tilde B (\la) \bigl(\tilde A (\m) - \tilde D (\m)\bigr) \epp
\end{align}
\end{subequations}
Here we referred back to the notation for the matrix elements
introduced in (\ref{defttilde}). Iterating (\ref{adb}) we obtain
\begin{multline} \label{ttb}
     \str_{g'} (\tilde \CT (\la))
        \Bigl[ \prod_{k=1}^M \tilde B(\m_k) \Bigr] =
     \Bigl[ \prod_{k=1}^M \tilde B(\m_k) \Bigr]
        \str_{g'} (\tilde \CT (\la))
	\prod_{k=1}^M \frac{1}{a(\la - \m_k)} \\
	- \sum_{k=1}^M \Bigl[ \tilde B(\la)
	  \prod_{\substack{l = 1 \\ l \ne k}}^M \tilde B(\m_l) \Bigr] 
	  s_{k-1} \str_{g'} (\tilde \CT (\m_k))
	  \frac{b(\la - \m_k)}{a(\la - \m_k)}
	  \prod_{\substack{l = 1 \\ l \ne k}}^M
	  \frac{1}{a(\m_k - \m_l)} \epc
\end{multline}
where the products over the $\tilde B (\m_k)$ are ordered in ascending
order and by definition
\begin{equation}
     s_{k-1} = \prod_{l=1}^{k-1}
               \bigl( b(\m_k - \m_l) - a(\m_k - \m_l) \bigr)
\end{equation}
for $k = 2, \dots, M$ and $s_0 = 1$. It follows that
\begin{equation} \label{evec2}
     \ph \, \Om = \tilde B(\m_1) \dots \tilde B (\m_M) \hat \Om
        \in \bigl( {\mathbb C}^2 \bigr)^{\otimes N} \otimes {\cal H}_0
\end{equation}
is an eigenvector of $\str_{g'} (\tilde \CT (\la))$ with eigenvalue
\begin{equation} \label{evtilde}
     \tilde \La (\la) =
        \Bigl[ a_2 (\la) - a_3 (\la) \prod_{j=1}^N a(\la_j - \la) \Bigr]
	\prod_{k=1}^M \frac{1}{a(\la - \m_k)}
\end{equation}
if the Bethe ansatz equations
\begin{equation}
     \frac{a_3 (\m_k)}{a_2 (\m_k)} =
        \prod_{j=1}^N \frac{1}{a(\la_j -\m_k)} \epc
	\qd k = 1, \dots, M \epc
\end{equation}
are satisfied. Inserting the expressions (\ref{evec2}) into
(\ref{evpre}) and (\ref{evtilde}) into (\ref{balpre}), (\ref{tevpre})
we arrive at the results shown in section \ref{sec:res} and our
derivation for the case $g = (-,-,+)$ is complete.

It is now relatively easy to perform the algebraic Bethe ansatz for the
remaining grading $(-,+,-)$. According to our remark at the beginning
of this appendix this case is equivalent to considering the
grading $(-,-,+)$ with the vacuum action shown in equation
(\ref{vacactmpm}). Therefore only the second level Bethe ansatz,
starting from equation (\ref{vacmmp}) has to be modified. Since
$D(\la)$ now acts as a lower triangular matrix on $\Om$ we must
choose the auxiliary vacuum for the second Bethe ansatz as $\hat \Om =
\binom{0}{1}^{\otimes N} \Om$. Then $\tilde \CT (\la)$ has lower
triangular action on $\hat \Om$, and an algebraic Bethe ansatz (with
$\tilde C$ replacing $\tilde B$) becomes again possible:
Introducing the explicit form of the $L$-matrix (\ref{lpm})
and the explicit form of ${\cal D_{(-+-)}}$, equation (\ref{dmpm}),
into the definition (\ref{defttilde}) of $\tilde \CT (\la)$ we obtain
\begin{equation}
     \tilde \CT (\la) \, \hat \Om =
        \begin{pmatrix}
	   a_3 (\la) \prod_{j=1}^N a(\la_j - \la) & 0 \\
	   \ast & a_2 (\la)
	   \prod_{j=1}^N \frac{a(\la_j - \la)}{a(\la - \la_j)}
        \end{pmatrix} \hat \Om \epc
\end{equation}
where we used the identity $a(\la) - b(\la) = a(\la)/a(-\la)$.
Instead of the commutation relations (\ref{tildecom}) we now need the
commutation relation between $\tilde C(\la)$ and $\tilde C(\m)$ and
between $\str_{g'} (\tilde \CT (\la))$ and $\tilde C(\m)$. These
commutation relations are again contained in (\ref{gyba11}). They are
of the same form as (\ref{tildecom}) with $\tilde C$ replacing
$\tilde B$ and the arguments $\la$, $\m$ of the functions $a$ and
$b$ interchanged. It follows that
\begin{equation} \label{evec2mpm}
     \ph \, \Om = \tilde C(\m_1) \dots \tilde C (\m_M) \hat \Om
        \in \bigl( {\mathbb C}^2 \bigr)^{\otimes N} \otimes {\cal H}_0
\end{equation}
is an eigenvector of $\str_{g'} (\tilde \CT (\la))$ with eigenvalue
\begin{equation} \label{evtildempm}
     \tilde \La (\la) =
        \Bigl[ a_3 (\la) \prod_{j=1}^N a(\la_j - \la) - a_2 (\la)
	\prod_{j=1}^N \frac{a(\la_j - \la)}{a(\la - \la_j)} \Bigr]
	\prod_{k=1}^M \frac{1}{a(\m_k - \la)}
\end{equation}
if the Bethe ansatz equations
\begin{equation}
     \frac{a_3 (\m_k)}{a_2 (\m_k)} =
        \prod_{j=1}^N \frac{1}{a(\m_k - \la_j)} \epc
	\qd k = 1, \dots, M \epc
\end{equation}
are satisfied. Inserting the expressions (\ref{evec2mpm}) into
(\ref{evpre}) and (\ref{evtildempm}) into (\ref{balpre}),
(\ref{tevpre}) we obtain the results for $g = (-,+,-)$ shown in
section \ref{sec:res}.

}

\newpage


\end{document}